\documentclass[aoas, preprint]{imsart}

\RequirePackage[OT1]{fontenc}
\RequirePackage{amsthm,amsmath,amssymb}
\RequirePackage[authoryear, colon]{natbib}
\RequirePackage[colorlinks,citecolor=blue,urlcolor=blue]{hyperref}
\RequirePackage{graphicx}
\RequirePackage[labelformat=simple]{subfig}

\startlocaldefs
\numberwithin{equation}{section}
\theoremstyle{plain}

\endlocaldefs

\newcommand{\argmax}{\operatornamewithlimits{argmax}}

\begin{document}

\begin{frontmatter}
\title{Survival-supervised latent Dirichlet allocation models
for genomic analysis of time-to-event outcomes}
\runtitle{survLDA}

\begin{aug}
\author{\fnms{John A.} \snm{Dawson}\thanksref{m1}\ead[label=e1]{jadawson@wisc.edu}}
\and
\author{\fnms{Christina} \snm{Kendziorski}\thanksref{t1,m2}\ead[label=e2]{kendzior@biostat.wisc.edu}}

\thankstext{t1}{Corresponding author}
\runauthor{Dawson and Kendziorski}

\affiliation{Department of Statistics, University of Wisconsin-Madison\thanksmark{m1} and
Department of Biostatistics and Medical Informatics, University of Wisconsin-Madison\thanksmark{m2}}

\address{
John A. Dawson\\
Department of Statistics\\
University of Wisconsin\\
\vspace{0.1in}
Madison, Wisconsin 53706\\
Christina Kendziorski\\
Department of Biostatistics and Medical Informatics\\
University of Wisconsin\\
Madison, Wisconsin 53706\\
Phone: 608-262-3146\\
\vspace{0.1in}
Fax: 608-265-7916\\
\printead{e1}\\
\phantom{E-mail:\ }\printead*{e2}}
\end{aug}

\begin{abstract}
Two challenging problems in the clinical study of cancer are the characterization of
cancer subtypes and the classification of individual patients according to those subtypes.
Statistical approaches addressing these problems are hampered by population heterogeneity
and challenges inherent in data integration across high-dimensional, diverse covariates.
We have developed a survival-supervised latent Dirichlet allocation (survLDA) modeling framework
to address these concerns. LDA models have proven extremely effective at identifying themes
common across large collections of text, but applications to genomics have been limited.
Our framework extends LDA to the genome by considering each patient as a ``document" with
``text" constructed from clinical and high-dimensional genomic measurements. We then
further extend the framework to allow for supervision by a time-to-event response.
The model enables the efficient identification of collections of
clinical and genomic features that co-occur within patient subgroups, and then characterizes
each patient by those features. An application of survLDA to The Cancer Genome Atlas (TCGA)
ovarian project identifies informative patient subgroups that are characterized by
different propensities for exhibiting abnormal mRNA expression and methylations,
corresponding to differential rates of survival from primary therapy.
\end{abstract}

\begin{keyword}
\kwd{Latent Dirichlet Allocation}
\kwd{Time-To-Event}
\kwd{Survival}
\kwd{Cancer}
\kwd{Genomics}
\end{keyword}

\end{frontmatter}

\section{Introduction}
Technological advances continue to increase both the ease and
accuracy with which measurements of the genome and phenome can be obtained
and, consequently, genomic-based studies of disease often
involve highly diverse types of data collected on large groups of patients.
The primary goals of such studies involve identifying
genomic features useful for characterizing patient subgroups as well as
predicting patient-specific disease course and/or likelihood of response to treatment.
Doing so requires statistical methods that handle complex interactions,
accommodate population heterogeneity, and allow for data integration across multiple sources.

A number of statistical methods are available for survival-related feature identification and
prediction (for a review, see \cite{Witten09, liLi2004, weiLi2007}).
Most often, classical models for a survival response are coupled with
some dimension-reduction method for
individual \citep{liLuan03, ghoshYuan10, pangDattaZhao2010}
or grouped predictors \citep{chenWang2009, liLi2004, ma07, ishwaran10},
providing a concise representation of the genomic features affecting patient outcome.
Although useful, the majority of these methods identify a set of covariates
common to all patients and as a result  may ``distort what is observed''
in the presence of population heterogeneity \citep{aalen88}.
Survival-supervised clustering approaches naturally accommodate heterogeneity, providing for
efficient and effective identification of patient subgroups
\citep{buhlmann02, liGui04}.  However, these approaches do not identify salient features
associated with subgroups; and, as with the aforementioned methods, may sacrifice power 
and accuracy by focusing on one (or a few) data sets in isolation.

Latent Dirichlet allocation (LDA, \cite{Blei03}) models are particularly well-tailored for accommodating
heterogeneity, selecting features, and characterizing complex interactions in a high-dimensional setting.
By far the most common application concerns identifying groups of words that co-occur frequently (topics) 
across large collections of text (e.g., a collection of abstracts, emails, or manuscripts).
The derived topics provide insight into the collections' content overall as well as into the specific
content within a document; and estimated document-specific distributions over topics 
are useful in classifying new documents (\cite{Blei03, Porteous08, Biro08}).

A recent extension allows for topic estimation to be supervised by a response that is suitably described
by a generalized linear model \citep{Blei08}. So-called supervised LDA (sLDA) debuted with a study of
movie reviews (text) and estimated topics (collections of co-occurring words in a review) that determined
the number of stars (supervising response) a movie received. Derived topics included ones 
having highest weight on words such as `power', `perfect', `fascinating' and `complex'; another 
with highest weight on `routine', `awful', `featuring', `dry'; a third on `unfortunately',
`least', `flat', `dull'; and so on. The movie-review-specific distribution over
topics proved useful in classifying movies. Those with highest weight on the
`power' topic generally had a high number of stars while those with highest weight on
the `unfortunately' topic had a low number; those with weight on the `routine' topic
most often ended up in the middle. 
Differences between the distributions also provided insights into differences 
between movies that received a similar number of stars, as well as insights
into the connotative nature of word choice (e.g., pictures described as `films' rating higher than
those referred to as `movies').

Our interest here is not in evaluating movies. However, it is important to note that
the questions addressed in \cite{Blei08} are identical in structure to the most important
questions we face in cancer genomics. In the former, questions include:
`Given reviews and ratings for a group of movies, can we identify 
topics - collections of words that co-occur frequently in some reviews and less frequently in others ?
Can each individual movie be described by a distribution over those topics ? Can distributions
over topics provide insights into differences between similarly rated movies ? 
And can a movie-specific distribution 
over topics be used to predict what the rating of a new movie will be ?'
In cancer genomics, the questions include: 
`Given genomic, clinical, and survival information on a group of patients, can we identify
topics that are collections of genomic and clinical features that co-occur frequently in some patients
and less frequently in others ? Can a patient be well described by a distribution over
those topics ? Can distributions over topics provide insights into the genomic 
differences 
between two patients with similar survival ? 
And can a patient-specific distribution over topics be used
to predict survival of an individual patient ?'

To address these types of questions, we extend
LDA for use in a clinical and genomic setting. Unlike in the textual domains of
\cite{Blei03}, \cite{Porteous08} and \cite{Biro08}, 
the definition of a document here is not obvious.  Section \ref{sec:cons} details the construction of documents, 
one for each patient, where words describe clinical events, treatment protocols, 
and genomic information from multiple sources. 
As we show in Section \ref{sec:app}, application of traditional LDA to this collection of documents provides for the identification of
topics useful in characterizing patient subpopulations as well as individual patients in a study of
ovarian cancer conducted as part of The Cancer Genome Atlas (TCGA) project \citep{tcga}. 
Survival supervised LDA (survLDA) is developed in Section \ref{sec:survLDA}
to facilitate topic supervision by a time-to-event response, 
which further improves patient-specific characterization and prediction.

\section{Methods}

\subsection{The LDA model.}\label{sec:lda}
We briefly review the LDA model as detailed in Blei et al. \citeyear{Blei03}. 
Assume there are $D$ documents indexed by $i = 1, \ldots, D$, each of which consists of $N_i$ words.
The vocabulary is the unique set of length $V$ indexed by $v = 1, \ldots, V$, from which the
documents' words arise, and is usually
taken to be the union of all words over documents. Further assume that there are $K$ latent `topics'
indexed by $k = 1, \ldots, K$, that govern the assignment of words to documents. Each topic corresponds to
a discrete distribution over the $V$ words in the vocabulary, with parameters given by the $V$-vector
$\tau_k$. Likewise, each document is assumed to be
associated with mixing coefficients $\theta_i$ over the $K$ topics, indicating its partiality with respect to word sources.

For a given document $i$, $N_i$ words arise from the following generative process
given the system-wide hyperparameters $\alpha$ (a $K$-vector Dirichlet parameter)
and the $\tau_{1:K}$ (the topic $V$-vectors):

\begin{enumerate}
\item Draw topic proportions $\theta_i \sim Dirichlet(\alpha)$
\item For each of the $N_i$ words, indexed by $j$:
\begin{enumerate}
\item Draw a topic assignment $Z_{ij} | \theta_i \sim Multinomial(1, \theta_i)$ \hspace{0.26in} (Note $Z_{ij} \in \{1, \ldots, K\}$)
\item Draw a word $W_{ij} | Z_{ij}, \tau_{1:K} \sim Multinomial(1, \tau_{Z_{ij}})$ \hspace{0.5in} (Note $W_{ij} \in \{1, \ldots, V\}$)
\end{enumerate}
\end{enumerate}

\noindent With this model in place, a variational expectation-maximization (EM) algorithm may be
used to estimate the joint posterior distribution
of $\theta_i$ and $Z_{i,1:N_i}$ given $w_{i,1:N_i}$, $\alpha$ and $\tau_{1:K}$
for each document $i$ (E-step) and then to estimate the system-wide hyperparameters
$\alpha$ and $\tau_{1:K}$ (M-step).
Upon convergence, the variational EM yields optimal values for the key
quantities of interest, namely posterior estimates for $\theta_{1:D}$ and $\tau_{1:K}$.

We note that the topics engendered by unsupervised LDA are, well, unsupervised. If there is a relationship
between certain words and an outcome of interest, the model may be hard pressed to find it
as the algorithm does not have access to the outcome. While one may certainly be interested in the structure
of documents and their constituent topics, the focus in our setting is often on
relationships between topics and an outcome of interest. When the outcome can be modeled by a 
generalized linear model, supervised LDA \emph{sLDA} can be used \cite{Blei08}.
Time-to-event outcomes such as survival, however, do not fall naturally into this class
due to the presence of (often abundant) censoring.
We have therefore extended the sLDA framework to time-to-event supervising outcomes,
a framework we will call \emph{survLDA}, in order to use these kinds of outcomes to inform the estimation
of topic distributions.

\subsection{The survival supervised LDA model} \label{sec:survLDA}

Assume the same setup as in Section \ref{sec:lda} with $D$ documents indexed by $i$, a vocabulary
of size $V$, and $K$ topics with corresponding discrete distributions $\tau_{1:K}$.
Additionally, the introduction of supervision through a time-to-event outcome means that,
just as a document's partiality to certain topics through $\theta_i$ impacts
its constituent words, those topics affect the survival outcome $T_i$. An indicator variable for death/censoring
is also observed for each document and denoted by $\delta_i$.

The system-wide model parameters for the survLDA model include a $K$-vector Dirichlet parameter
$\alpha$, the topic $V$-vectors $\tau_{1:K}$, and survival response parameters $\beta$ (a $K$-vector of regression coefficients)
and $h_0(\cdot)$ (baseline hazard). When they are available for a given document $i$,
$N_i$ words and a survival response $T_i$ arise from the following generative process:

\begin{enumerate}
\item Draw topic proportions $\theta_i \sim Dirichlet(\alpha)$
\item For each of the $N_i$ words, indexed by $j$
\begin{enumerate}
\item Draw a topic assignment $Z_{ij} | \theta_i \sim Multinomial(1, \theta_i)$ \hspace{0.26in} (Note $Z_{ij} \in \{1, \ldots, K\}$)
\item Draw a word $W_{ij} | Z_{ij}, \tau_{1:K} \sim Multinomial(1, \tau_{Z_{ij}})$ \hspace{0.5in} (Note $W_{ij} \in \{1, \ldots, V\}$)
\end{enumerate}
\item Compute the $K$-vector $\bar{Z}_i$ s.t. $\bar{Z}_{ik} = \#\{Z_{ij} = k\} / N_i$
\item Draw a survival response $T_i | \bar{Z}_i, \beta, h_0$ from the survival function corresponding to
a Cox proportional hazards model with hazard function $h(t | \bar{Z}_i) = h_0(t) \exp\{\beta^\prime \bar{Z}_i\}$
\end{enumerate}

\noindent Note that we are using a Cox proportional hazards model \citep{Cox72},
with each regression coefficient $\beta_k$ exhibiting the beneficent (negative) or deleterious (positive)
effect of topic $k$ on survival. The form of $h_0$ may be chosen by the user and may be parametric
(such as using a Weibull survival model) or nonparametric. 

As in LDA, a variational EM algorithm may be
used to estimate the joint posterior distribution
of $\theta_i$ and $Z_{i,1:N_i}$ given $w_{i,1:N_i}$, $T_i$, $\delta_i$, $\alpha$, $\tau_{1:K}$, $\beta$ and $h_0$
for each document $i$ (E-step) and then to estimate the system-wide hyperparameters
$\alpha$, $\tau_{1:K}$, $\beta$ and $h_0$ (M-step). 
The derivation of this variational EM is given in Appendix A. 
As detailed in Appendix B, we introduce into the variational EM an `uninteresting' background topic to
act as a benchmark with respect to the supervising outcome.
Without loss of generality, say this is the last ($K^{th}$) topic. 
Then the distribution over the vocabulary for the background topic
($\tau_K$) may be identified by placing count weights on the known `background' words, a tiny
amount of weight on all other words (say, 0.001) and re-normalizing so that the sum over the
entire $V$-vector is one.  Non-background topics are then different deviations
from this benchmark that express themselves through differential survival.
In our application, the background topic would describe `featureless' documents that
contain only the ubiquitous adjuvant therapy information, nothing more.
Upon convergence, the variational EM yields posterior estimates for the key
quantities of interest: posterior estimates for the $\theta_{1:D}$ as well as for
the composition ($\tau_{1:K}$) and outcome effect ($\beta$) of the $K$ topics.

\subsection{Prediction} \label{sec:prediction}

Given a new patient with document $w_{1:N}$ and a fitted model $\{\alpha, \tau_{1:K}\}$, the posterior mean
$\bar{Z}_{new} = \bar{Z} | w_{1:N}, \alpha, \tau_{1:K}$ can be obtained in order
to estimate from what topics this new patient draws words, and in what proportions.
As was the case during model fitting, this posterior must be approximated via variational inference.
We do so by following the same procedure as outlined in the first subsection of Appendix A,
except that all survival-related terms in the evidence lower bound are dropped; 
see the third subsection of Appendix A for
details. We note that this approach is analogous to that in \cite{Blei08},
where they point out that the prediction protocol does not depend on the particular response type.

Given $\bar{Z}_{new}$, measures related to topic membership can be predicted for the new patient.
This may be done qualitatively (e.g., ``This patient is predicted to belong strongly to the second topic
and survival for that topic is poor, hence her prognosis is bad.") or quantitatively (e.g., predicting
median survival time using the parametric survival model; see Appendix A).

\subsection{Document construction in the TCGA cohort} \label{sec:cons}

In most applications of LDA, the definition of `document' is obvious. That is not the case here.
In studies of cancer genomics, data are available in the form of text (e.g., treatments received, 
a clinician's evaluation of response to treatment, etc.); there are also often
disparate non-textual measurements (e.g., binary, count, continuous, factor).
By translating these measurements into words, and thus patients into textual documents, 
the LDA and survLDA models may be used for inference.

Our population of interest is the ovarian cancer cohort from The Cancer Genome Atlas (TCGA) \citep{tcga}.
For these women, clinical information such as time of surgery, adjuvant
therapies, time of recurrence, treatment at recurrence, overall survival, and dozens of other variables
are available. Also available are high-throughput measurements of gene expression, methylation, 
SNP/CNVs, and microRNAs.  For document construction, we used words associated with adjuvant therapy, 
expression as assayed by the Affymetrix HT Human Genome U133A chip, and methylation as assayed using
the Illumina Infinium HumanMethylation27 BeadChip.  
Out of the cohort as accessed in April of 2011, 448 of the women
were not missing more than half of their gene expression or methylation measurements
and received at least the standard therapy of a platinum and taxane.
A document was constructed for each of the 448 patients.

For a given patient, a drug-related word was added to her document for every drug given to her during adjuvant treatment.
For example, a patient receiving two platinums and a taxane would have the words `platinum', `platinum' and `taxane'
added to her document.
Related drugs were collected together (e.g., the many varied platinums) and 
adjuvant drug therapies not given to at least 10\% of the women were not considered.  
To help ensure that documents contained words corresponding to meaningful genes, we considered 
those genes for which expression or methylation is multi-modal, as assessed by MCLUST 
\citep{FraleyRaftery2002, MclustSoftwareManual2006}. Specifically, a gene's expression distribution was
deemed multi-modal if MCLUST preferred a two-component over a one-component model when given only those choices.
Filtering by known genes, non-missingness, and multi-modality reduced the number of gene expression measurements
considered from 14,500 to 7,727. 
Given the multi-modal genes, a patient's document received a gene word, given by the gene name, 
if that patient showed extreme expression for that gene.  
The same word was added again if the patient showed extreme methylation.
As most genes have multiple sites at which methylation was measured, we considered each methylation
site in each gene separately. So a patient's document receives copies of the word
``geneX'' if that patient had extreme expression for geneX (one copy added) 
as well as multiple hyper (or hypo) methylated sites (one copy added per hyper- or hypo-methylated site).  
For both types of data, extreme was defined as being in the minor mode of the genes' multi-modal distribution,
if the probability of being so, as estimated by MCLUST, exceeded 95\%.
As extreme may manifest as high or low expression 
(or methylation), one cannot infer direction of the gene expression or methylation measurement from
word frequency; i.e., high frequency does not necessarily correspond to high expression.  
This is a consequence of the way in which documents were constructed and could be changed.

In an effort to improve power to detect survival-related topics, a second filter was applied.
Specifically, we considered the 234 uncensored patients and partitioned them into three groups of 
78 based on survival times.
A gene's word was removed if word frequency did not vary across groups. For gene expression (methylation), we required 
a difference of at least 10 (15) words between at least two of the survival groupings, leaving
201 (1,063) vocabulary words. Note that the censored women still received words based on
gene expression and methylations; they merely did not contribute to the decisions made during this pre-filtering.
A typical document contained approximately 350 words (mean 347; median = 132), on the order of a
PubMed abstract.

\section{Application to TCGA data} \label{sec:app}

Given documents constructed as described above for each of the 448 women considered, we applied 
LDA as well as survLDA.
The outcome of interest is all-cause mortality and for the survLDA application, we used a Weibull model 
for the baseline hazard.  The background topic (the seventh topic) was assigned 
non-trivial weight only on the adjuvant therapy words
platinum and taxane, as this setup would constitute an `uninteresting' document since all patients received these
two treatments (they were inclusion criteria in the TCGA ovarian project).  
In all analyses, we use $K=7$ topics, the last being the background topic.

\subsection{Results} \label{sec:appTCGA}

\begin{center}
\begin{figure}[LDAtopics]
  \centering
  \includegraphics[width=\textwidth, trim=0 0 100 0]{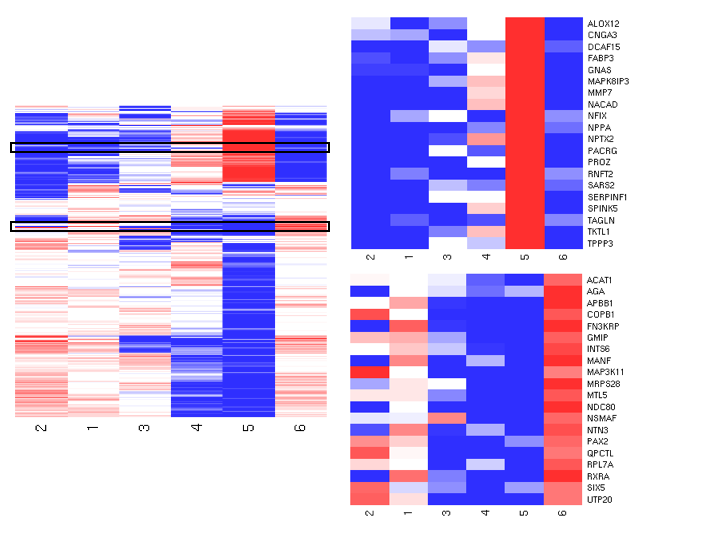}
  \caption{A heat map of the word distributions for the topics derived from LDA,
  with two insets marked by black boxes. The background (seventh) topic is not shown.
  Words are clustered along the rows, topics are clustered
  across the columns. The colors range from blue (word under-represented in the topic)
  to red (word over-represented in the topic), with white in the middle (average representation).
  The insets are close-up views of the boxed regions, which contain the twenty highest-weight words
  for topics five (upper right heatmap) and six (lower right heatmap).}
  \label{fig:LDAtopics}
\end{figure}
\end{center}

Application of (unsupervised) LDA provides two quantities of primary interest.
The first are the topics $\tau_{1:K}$, or estimated distributions over words; and the second are
the document-specific distributions over topics $\theta_{1:D}$.
The left panel of Figure \ref{fig:LDAtopics} presents
the topic-specific distributions over words for each topic.
Red (blue) indicates an overabundance (dearth)
of a word's weight in the corpus belonging to a particular topic;
white indicates an amount equivalent to an even spread over all non-background topics.
The overabundance of some words and paucity of others 
characterize the topics, allowing one to differentiate among them. 
For example, consider topics 5 and 6. The right panel of Figure \ref{fig:LDAtopics} 
shows twenty words having high weight
in topic 5 (upper), and twenty with high weight in topic 6 (lower). 
It is clear that patients described primarily by these topics differ with respect to aberrations
for this collection of genes. Of interest is determining whether this difference translates to a 
difference in overall survival.

\begin{center}
\begin{figure}
  \centering
  \begin{tabular}{cc}
  \subfloat[]{\includegraphics[width=0.45\textwidth, clip=true, trim=100 30 30 80]{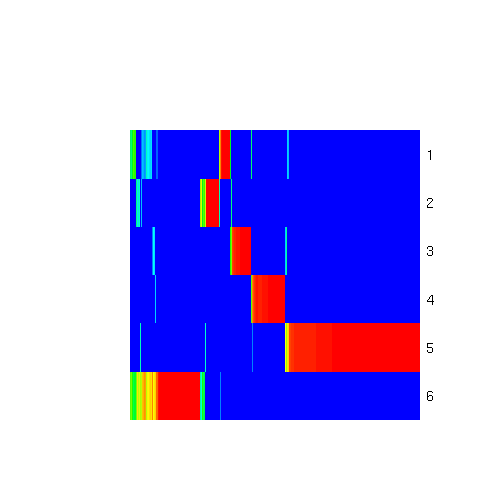}} &
  \subfloat[]{\includegraphics[width=0.5\textwidth, clip=true, trim=0 30 20 0]{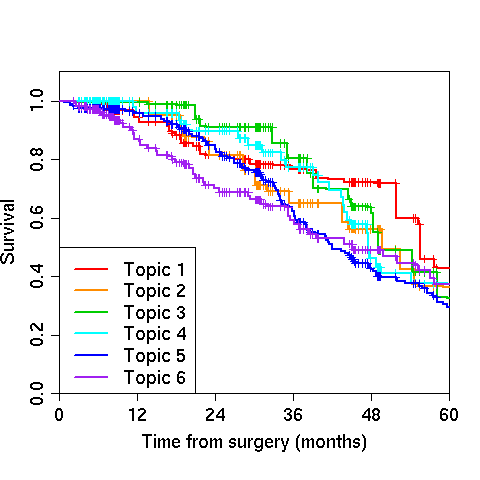}} \\
  \subfloat[]{\includegraphics[width=0.45\textwidth, clip=true, trim=100 30 30 80]{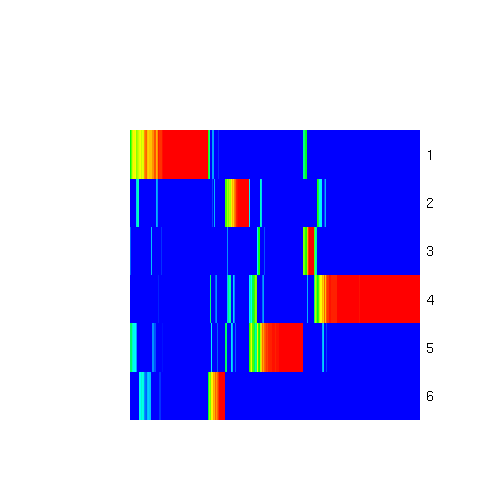}} &
  \subfloat[]{\includegraphics[width=0.5\textwidth, clip=true, trim=0 30 20 0]{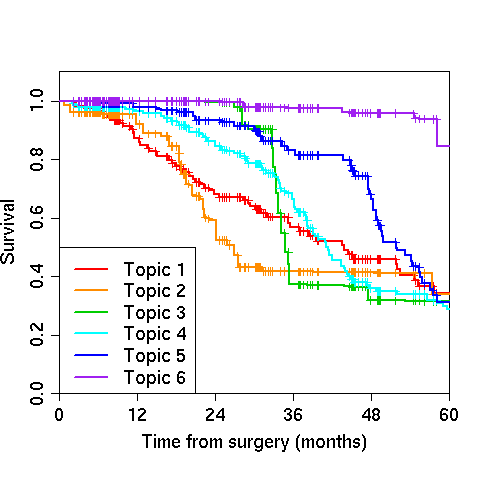}}
  \end{tabular}
\caption{Top left: A heat map of estimates of $\theta$ (patient-specific distributions over topics) 
for each patient derived from LDA (upper left) and survLDA (lower left).
  Topics are given across the rows and patients are clustered
  across the columns. Colors range from deep blue (0) to red (1). The right panels show Kaplan-Meier curves for the
  LDA (upper) and survLDA (lower) topics.  The background (seventh) topic is not shown. 
Topic $k$'s curve is generated using all 448
  documents, weighted by the $\theta_{ik}$.}
\label{fig:fourpanels}
\end{figure}
\end{center}

The top left panel of Figure \ref{fig:fourpanels} shows a heat map of estimated topic membership
for the six non-background topics (rows), clustered over patients (columns). 
As expected, none of the patients exhibited more than minimal weight in the background topic (not shown).
Patient membership within a topic ranges from 0 (nil weight in the topic, deep blue) 
to 1 (wholly belonging to the topic, red). As shown,
most patients have high weight in a single topic, while a few are best described by mixtures over topics.

Topics 5 and 6 are the largest topics, in the sense that they contain the most weight over the cohort.
The top right panel of Figure \ref{fig:fourpanels} 
shows Kaplan-Meier (KM) curves for each topic shown in the left panel.
The curves are generated by weighting the TCGA patients' survival information by their topic membership;
hence a patient whose $\theta_i$ had 50\% weight in topic one and 50\% weight in topic
two would count as `half a person' in the KM estimation of those curves, but would not contribute to
the survival curves associated with other topics.  As shown,
there is some, albeit limited, separation with respect to survival over topics.
Topics 5 and 6, for example, have rather similar survival (80\% vs 70\% at two years), in spite of the
genomic differences highlighted in Figure \ref{fig:LDAtopics}.
This type of finding is consistent 
with recent studies showing remarkable genetic heterogeneity among cancer patients that
appear to be clinically similar \citep{jones08}. 

The bottom left and right panels of Figure \ref{fig:fourpanels} similarly show estimated topic membership
and KM survival curves for the survLDA analysis. 
While the analysis uses a parametric Weibull survival model, empirical (weighted) KM curves
are presented to facilitate comparison to the (unsupervised) LDA approach. 
In survLDA, the largest topics are 1 and 4;
and better separation among all topics is observed.
Two topics have poorer than baseline survival (topics 1 and 2), 
and two have better (topics 5 and 6).

\begin{center}
\begin{figure}[survLDAtopics]
  \centering
  \includegraphics[width=\textwidth, trim=0 0 100 0]{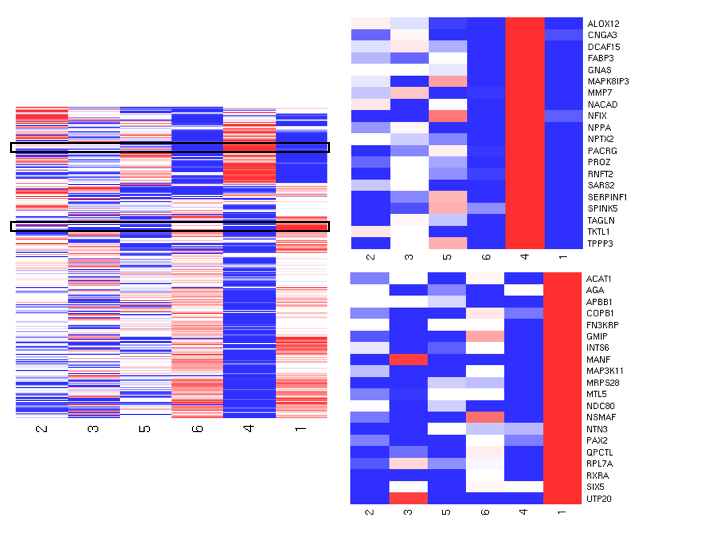}
  \caption{A heat map of the word distributions for the topics derived from survLDA,
  with two insets marked by black boxes. The background (seventh) topic is not shown.
  The rows are in the same order as in Figure 2, topics are clustered across the columns.
  The colors range from blue (word under-represented in the topic)
  to red (word over-represented in the topic), with white in the middle (average representation).
  The insets are close-up views of the boxed regions, which contain the twenty highest-weight words
  for topics 5 and 6. 
  Note the similarities between topics 4 and 1 derived from survLDA and topics 5 and 6
derived from LDA.}
  \label{fig:survLDAtopics}
\end{figure}
\end{center}

Figure \ref{fig:survLDAtopics} is similar to Figure \ref{fig:LDAtopics}, 
but shows results from survLDA. For comparison, the word order
shown in Figure \ref{fig:LDAtopics} is preserved in Figure \ref{fig:survLDAtopics}. 
Topics have also been reordered in Figure \ref{fig:survLDAtopics} to
stress similarity between topics 5 and 6 derived from the LDA analysis and topics 4 and 1 derived from survLDA.
As shown in Figure \ref{fig:survLDAtopics}, their topic-specific distributions over words are very similar.
This is not an artifact. Indeed, many of the topics' high-weight words are consistent between LDA and survLDA.
The consistency as we shift from free-formed topics to survival supervision suggests that there are subpopulations
within the cohort whose constituent topics have strong enough effects on survival
that they are evident even without survival supervision. When survival information is available 
and used to supervise topic creation, the subsequent alterations to topics 
result in a much sharper disparity in survival rates.
Recall that two year survival is about 80\% vs. 70\% for topics 5 and 6 in the LDA 
analysis; that spread increases to 92\% vs. 65\% under survival supervision for topics 
4 and 1 (which are most similar in structure to topics 5 and 6). 
We note that this magnitude of differential survival is on par with results
of other ovarian cancer studies \citep{tothill08, tcgaNature}.

\begin{center}
\begin{figure}[fig:wordclouds]
  \centering
  \includegraphics[width=\textwidth, trim=100 100 100 100]{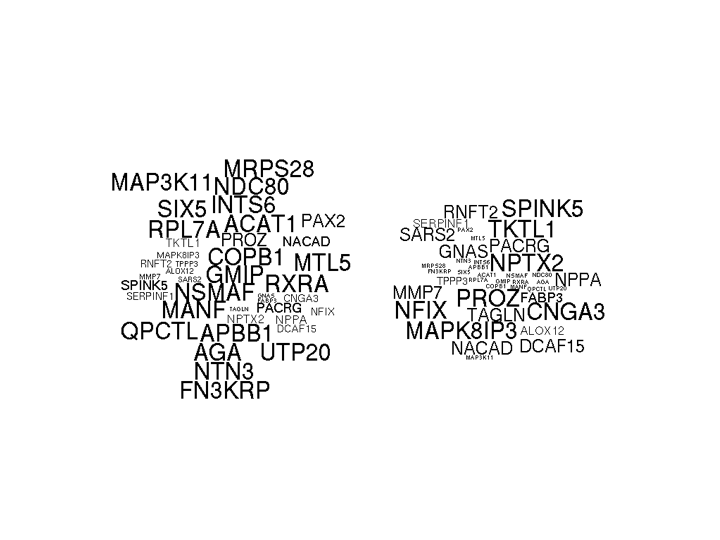}
\caption{Word clouds generated using the twenty highest-weight words 
for topics 1 (left) and 4 (right) derived from survLDA.}
\label{fig:wordclouds}
\end{figure}
\end{center}

With respect to words that distinguish between topics 4 and 1, we
note that the following words consistently show high frequency in the poor survival group:
NDC80 (a spindle checkpoint regulator associated with breast cancer \citep{ndc80})
RXRA (a transcriptional regulator associated with
breast cancer \citep{rxraRef1, rxraRef2}), MANF (a gene coding a highly conserved protein with unknown function with
mutations often observed in lung, breast, prostate \citep{manfRef1} and pancreatic cancers \citep{manfRef2}),
and INTS6 (a tumor suppressor known to be involved in prostate cancer \citep{ints6}).
Further note that NDC80, RXRA and INTS6 would not likely have been identified in this cohort
by another approach, as the marginal p-values from a Cox proportional hazards test
are far from overwhelming (NDC80 p=0.6, 0.98, 0.07; RXRA p=0.22, 0.37, 0.27;
INTS6 p=0.38, 0.8, 0.59; all p-values are for gene expression and then for two methylations
measured in the gene). MANF (p=0.02, 0.46, 0.08) is the only one of these that shows
even a nominally significant marginal effect on survival.
Further investigation of these and other genes that display markedly different abundance patterns between
these patient subtypes might elucidate the mechanisms that underlie differences between the groups.
To this end, word cloud representations such as those shown in Figure \ref{fig:wordclouds} may prove useful.

\begin{center}
\begin{figure}[barcodey]
  \centering
  \includegraphics[width=0.7\textwidth, trim=80 30 100 80]{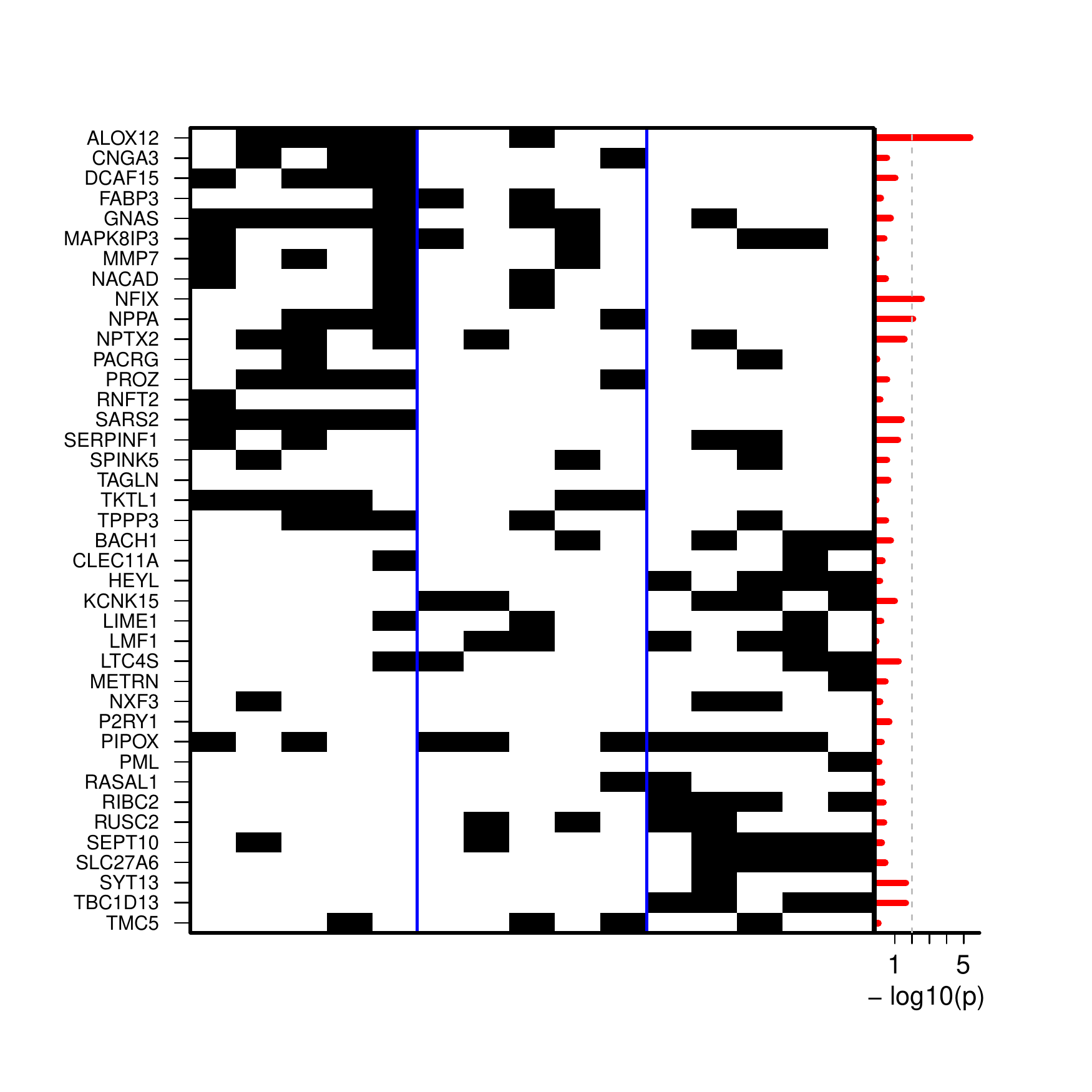}
  \caption{The rows show the twenty highest-weight words from the fourth and fifth topics derived from survLDA.
  The columns show words (presence in black and absence in white)
  for fifteen patients: 5 identified as strongly belonging to topic 4 
  $(\theta_{i,4} > 0.99$, left column$)$, 
5 identified as strongly belonging to topic 5 $(\theta_{i,5} > 0.99$, right column$)$,
  and 5 for whom the model indicates a mixture of the two
  $(\theta_{i,5} > 0.4$ and $\theta_{i,5} > 0.4$, middle column$)$. The
  bars on the right indicate the marginal p-values for each gene, which is taken to be the minimum of all p-values
  from Cox proportional hazard-based score tests among all expression and methylation measurements associated with
  that gene. Note that only three of the genes shown would have survived a pre-filtering step
  using a nominal 0.01 level.}
  \label{fig:barcode}
\end{figure}
\end{center}

Additionally, the patient-specific distributions over topics 
are useful in that they characterize the genomic aberrations
underlying individual patients.  Figure \ref{fig:barcode} shows 20 high-weight words from topic
4 and 20 from topic 5 (rows). Fifteen patients are shown (columns):
five identified as strongly belonging to topic four
$(\theta_{i,4} > 0.99$, left column$)$, five identified as strongly belonging to 
topic five $(\theta_{i,5} > 0.99$, right column$)$,
and five whom the model classifies 
as a mixture of both $(\theta_{i,5} > 0.4$ and $\theta_{i,5} > 0.4$, middle column$)$.
(Note that topics 4 and 5 were chosen since most patients are described almost exclusively by a single topic
as shown in Figure \ref{fig:fourpanels};
however, there are some best described by a mixture. In the survLDA analysis, most of these latter patients 
are a mixture between topics 4 and 5.)

The differences in genomic aberrations shown among the groups in Figure \ref{fig:barcode} are clear. 
Patients defined primarily by 
topic 4 have many of the high-weight topic 4 words in their document. The same is true for topic 5,
while those who are a mixture have some realizations from both sets of words. 
The right side of the plot shows p-values from Cox proportional hazards tests conducted on the
entire cohort. For each gene, a test
was conducted for expression as well as methylation measurements associated with that gene.
The p-value reported is the minimum among those tests. As shown, few of these genes 
would be identified as significant
using a standard Cox based test since the gene expression measurements are not significantly associated
with survival across the entire cohort, even though they show clear differences for some subsets of patients.

To evaluate the utility of survLDA for patient-specific prediction, we split the TCGA cohort into
a training and test set (75\% and 25\% of the cohort, respectively).
The full barcoding, document creation and survLDA model fitting procedures were applied
to the training set. Documents for the test set women were derived using
the abnormality indications for the genes and methylations surviving filtering
that arose from the training set document creation.
These documents in hand, the survLDA output was used
to predict topic membership for the test set, using the prediction approach
given in Section \ref{sec:prediction}.

\begin{center}
\begin{figure}[predFig]
  \centering
  \includegraphics[width=\textwidth]{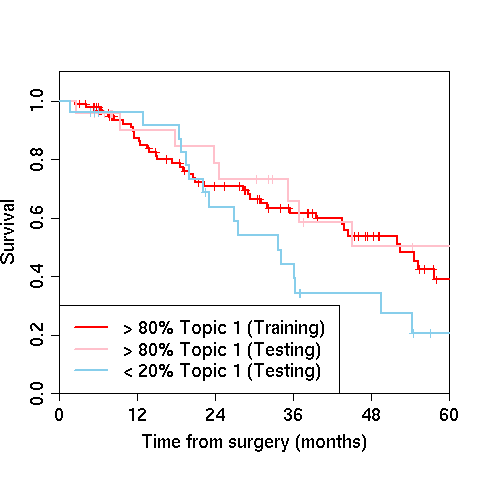}
  \caption{Kaplan Meier curves for selected topic structures within the training and testing data.
  The red (pink) line represents the 99 training set (23 testing set) women whose documents
  are identified as having more than 80\% weight in topic 1;
  the light blue line is the survival curve for the 26 test set women
  predicted as having less than 20\% weight in topic 1.}
  \label{fig:predFig}
\end{figure}
\end{center}

There were 23 women in the test set with more than 80\% weight in topic 1.
The patient-specific distributions over topics for the other women in the test set 
were not concentrated on a different topic,
but had weight rather evenly spread out across topics 2-6.
The pink (red) line in Figure \ref{fig:predFig} 
shows the Kaplan-Meier curve for the test (training) set women predicted as having more than 80\% weight
in topic 1, while the light blue line is the survival curve for test set women
predicted as having less than 20\% weight in topic 1.
As shown in Figure \ref{fig:predFig}, survival of women largely described by topic 1 in the training and test sets
is similar, while survival for those test cases predicted as not being well described by topic 1  
is considerably different, indicating some predictive power
for patients strongly described by a single topic.  Simulations
(not shown) indicate that topic specificity as well as 
prediction improves with either more or larger documents; and work toward these ends is ongoing.

\section{Discussion} \label{sec:disc}

A problem pervasive in genomic based studies of disease concerns
taking large, diverse data sets collected on a cohort of patients
and using the information contained therein to characterize patient subtypes as well
as individuals.  Computational scientists often address this problem by performing
analysis within a single data type and comparing results subsequently
in an effort to identify a signal supported by the disparate analyses (e.g., a gene's SNPs, expression,
and methylation all associate with a phenotype).
Comparing results manually has its obvious disadvantages. At the same time,
meta-analysis approaches such as Fisher's combined probability test can be
limited by low power \citep{zaykin02}; and efforts to combine data directly
are challenged by measurements on different scales with differential dependencies.
The LDA based framework proposed here addresses these challenges
by transforming the information contained in high-throughput genomic screens into text. 
Doing so has both advantages and disadvantages.

One advantage is that data integration is seamless. In the implementation presented, a  
word for a gene is assigned to a patient's document if the gene shows extreme expression;
the same word is assigned if the gene shows extreme methylation. In this way, a document may
contain copies of words associated with extreme genomic features, measured from expression
and/or methylation. The number of copies is proportional to the number of measurements for which
the gene is extreme: a gene with extreme expression and methylation will have more copies of that word
than a gene showing extreme expression alone. Although we have demonstrated the approach
using expression and methylation measurements (and treatment information), applications that use additional
types of data are easily incorporated into the framework.

A second advantage is that the threshold required for a gene to be included in the analysis
is much lower than would be required with other methods. As detailed in Section \ref{sec:cons}, some pre-selection of
genes is done, but the selection does not require even nominally significant association with a survival end-point,
as is often required in survival studies with high-dimensional covariates \citep{liLuan03, chenWang2009, liuTan09}.
As shown in Figure \ref{fig:barcode}, this allows for the identification of many important genes, 
some previously known to be involved in cancer, that would not otherwise have been considered.

Application of LDA to patient documents reveals groups
of genomic aberrations that co-occur together (topics) and then characterizes individual
patients by those groups. The topics themselves are useful in that they define collections of
genes, methylations or other covariates among which undiscovered interactions might occur, 
while the patient-specific distributions over topics give insights into the similarities
and differences among patients that go beyond the information that can be gained from
grouping by like outcome. In our application
to the TCGA ovarian project, there was some consistency between the topics
derived under unsupervised and supervised analyses, suggesting
that the approach may produce topics whose constituents have similar survival profiles
even in settings where survival information is not available. Of course, when survival is available,
survLDA may be used to improve inference.

Our analysis of the TCGA ovarian cohort identified several genes whose products and
methylations bear further interrogation, some already known to be involved in cancer.
Our investigations of the predictive ability of the approach are mixed. While
there is some ability for prediction, improvements are expected with
increases in the number of patients as well as improvements in document creation strategies.
For instance, in the testing set we could evaluate predictive inference in one group,
the only one for which the test women had strong indications of topic membership or exclusion.
Simulations (not shown) suggest that this limitation is largely due to the relatively small sample size
in the test set and the minimal amount of word
replication in our documents. More work is 
required to ascertain the power associated with sample size, document size, word frequencies, and
replication, which in turn requires further experimentation with the method of document construction.

As this framework relies heavily on the words contained in a patient's document, 
much work is required to develop and evaluate methods of document construction. 
Our approach to assign a word for any gene showing extreme expression or methylation
was motivated by \cite{zilliox07}, where
the authors identify bi-modal genes and, 
for each individual and each gene, assign a binary variable indicative of 
mode membership.  The resulting gene expression `barcode' for each patient proved extremely useful in 
classifying patients into biologically meaningful groups in \cite{zilliox07}; and, as demonstrated,
the extrapolation of their approach proved to be an effective strategy here. 
At the same time, one could imagine assigning
a word associated with a pathway if any gene in that pathway was extreme as assessed by 
expression, methylation, SNP profile, etc.   This would increase the frequency with which
words appeared; and our preliminary evaluation suggests that this can be useful, but can also
result in topics that are not clearly distinct with respect to high-weight words.
Another possibility is to assign an increasing number of words in direct 
proportion with signal.  For example, consider breaking a gene's expression into deciles, say, and
assign one to ten words for
each document (e.g., a value between the sixth and seventh deciles gets seven words).
We did not favor this approach for two main reasons. First, the approach assumes linearity of 
expression and methylation which is often not the case. 
Second, the approach results in documents having few unique words, which reduces specificity of
topics as well as document specific distributions over topics. 
Document construction continues to be explored, and improvements are expected to prove useful
in a number of settings.

In addition to the means by which covariates are translated into words, there are many aspects
of the proposed methods that require further development.
In particular, survLDA assumes the simplest of Dirichlet priors on the distributions of topics over
patients and therefore the documents are considered conditionally independent given $\alpha$.
While this is a reasonable assumption for the TCGA data set we considered, there
are other realms where correlation among the documents could arise. For example,
one could have multiple documents arising from the same subject, one for each time point or tissue;
or, when integrating multiple cancer types, subjects with the same type of cancer
would be expected to be more alike than subjects with differing cancer types. Adding such hierarchy
has already been explored for traditional LDA \citep{Teh06}, presenting
a starting point for methodological extension.

Similarly, the composition of the topics themselves is essentially free. Were it not for our
imposition of a background topic, the topics would be completely unstructured {\it a priori}.
As it is, $K-1$ topics are still governed solely by the data. This need not be the case,
as methods similar to those proposed for construction of a background topic (see Appendix B) could
be extended. In particular,
the Dirichlet prior could be modified directly, or a set of restrictions could be
imposed for each topic and groups of words
so that certain words cannot appear together, or may only appear together in certain topics.
Some of these modifications were considered in
\cite{AndrzejewskiZhu09}, providing another starting point for future extension.

In summary, it is becoming increasingly clear that studies aimed at solving 
the most challenging problems in cancer genomics involve highly diverse types of data
collected on large groups of patients. Many methods will prove useful. We suspect that
advantage will be gained from methods that are able to integrate data and account for cohort heterogeneity,
allow supervision by outcomes of interest such as survival, provide for patient specific inference,
and facilitate prediction of unobserved outcomes. 
The proposed approach provides tools for these purposes in an effort to help
ensure that maximal information is obtained from powerful genomic based studies of disease.

\appendix

\section*{Appendix A: The survLDA variational EM}

\subsection*{Posterior inference} \label{sec:postInf}

For a given document $i$ with survival response dyad $(T_i, \delta_i)$, the key quantity of interest is

\begin{flalign}
p(\theta_i, Z_{i,1:N_i} | w_{i,1:N_i}, T_i, \delta_i, \alpha, \tau_{1:K}, \beta, h_0) &=& \nonumber
\end{flalign}
\begin{flalign}
&&\frac{p(\theta_i | \alpha) \left(\prod_{j=1}^{N_j} p(Z_{ij} | \theta_i) p(W_{ij} | Z_{ij}, \tau_{1:K}) \right)
p(T_i, \delta_i | Z_{i,1:N_i}, \beta, h_0)}
{\int \, p(\theta_i | \alpha) \sum_{Z_{i,1:N_i}} \left(\prod_{j=1}^{N_i} p(Z_{ij} | \theta_i)
p(W_{ij} | Z_{ij}, \tau_{1:K})\right) p(T_i, \delta_i | Z_{i,1:N_i}, \beta, h_0) \, \mathrm{d}\theta} \label{eq:postdoc}
\end{flalign}

\noindent where the normalizing value is known as the \emph{evidence}.
As in LDA \citep{Blei03} and sLDA \citep{Blei08}, the evidence cannot be exactly computed efficiently,
so we will use mean-field variational inference using Jensen's inequality to approximate it. For reviews of this
and other variational methods, see \cite{Wainwright08} and \cite{Jordan99}.

Let $\pi = \{\alpha, \tau_{1:K}, \beta, h_0\}$ and $q_i(\theta_i, Z_{i,1:N_i})$ denote a \emph{variational distribution}
of the latent variables. For computational tractability, we choose a fully factorized variational distribution:
\begin{align}
q_i(\theta_i, Z_{i,1:N_i} | \gamma_i, \phi_{1,1:N_i}) = q_i(\theta_i | \gamma_i) \prod_{j=1}^{N_i} q_i(Z_{ij} | \phi_{ij}) \label{eq:vardist}
\end{align}
\noindent where
\begin{eqnarray}
\theta_i | \gamma_i \sim Dir(\gamma_i) \hspace{0.5in} \text{and} \hspace{0.5in} Z_{ij} | \phi_{ij} \sim Discrete(\phi_{ij}). \nonumber
\end{eqnarray}

\noindent With this quantity defined, the lower bound for the evidence given by Jensen's inequality is
\begin{eqnarray}
\log \, p(W_{i,1:N_i}, T_i, \delta_i | \pi) &=& \log \int_{\theta_i} \sum_{Z_{i,1:N_i}} p(\theta_i, Z_{i,1:N_i}, W_{i,1:N_i}, T_i,
\delta_i | \pi) \, \mathrm{d}\theta \nonumber \\
 &=& \log \int_{\theta_i} \sum_{Z_{i,1:N_i}} p(\theta_i, Z_{i,1:N_i},
 W_{i,1:N_i}, T_i, \delta_i | \pi) \frac{q(\theta_i, Z_{i,1:N_i})}{q(\theta_i, Z_{i,1:N_i})} \, \mathrm{d}\theta \nonumber \\
 &=& \log \, E_{q_i} \left[p(\theta_i, Z_{i,1:N_i}, W_{i,1:N_i}, T_i, \delta_i | \pi) \frac{1}{q(\theta_i, Z_{i,1:N_i})} \right] \nonumber \\
 &\geq& E_{q_i} \left[ \log \, p(\theta_i, Z_{i,1:N_i}, W_{i,1:N_i}, T_i, \delta_i | \pi) \frac{1}{q(\theta_i, Z_{i,1:N_i})} \right] \nonumber \\
 &=& E_{q_i}[\log \, p(\theta_i, Z_{i,1:N_i}, W_{i,1:N_i}, T_i, \delta_i | \pi)] + -E_{q_i}[\log \, q(\theta_i, Z_{i,1:N_i})] \label{eq:elbo}
\end{eqnarray}
\noindent where the second term in the lower bound is the entropy $H(q_i)$ of the variational distribution.
We will use $\mathcal{L}(\cdot)$ to refer to the so-called \emph{evidence lower bound} (ELBO)
given in (\ref{eq:elbo}). We can expand the ELBO:
\begin{eqnarray}
\mathcal{L}(W_{i,1:N_i}, T_i, \delta_i | \pi) &=& E_{q_i}[\log \, p(\theta_i | \alpha)]
+ \sum_{j=1}^{N_i} E_{q_i}[\log \, p(Z_{ij} | \theta_i)] + \sum_{j=1}^{N_i} E_{q_i}[\log \, p(W_{ij} | Z_{ij}, \tau_{1:K})]
\nonumber \\ & & + E_{q_i}[\log \, p(T_i, \delta_i | Z_{i,1:N_i}, \beta, h_0)] + H(q_i) \label{eq:exelbo}
\end{eqnarray}

\noindent Thus, an approximation of the posterior given in (\ref{eq:postdoc}) is obtained by maximizing $\mathcal{L}$ with
respect to $\gamma_i$ and $\phi_{i,1:N_i}$. The first, second and third terms in (\ref{eq:exelbo}), as well as the entropy $H(q_i)$,
are identical to the corresponding terms in the ELBO for LDA \citep{Blei03} and sLDA \citep{Blei08}:

\begin{eqnarray}
E_{q_i}[\log \, p(\theta_i | \alpha)] &=&
\log \, \Gamma \left( \sum_{k=1}^K \alpha_k \right) - \sum_{k=1}^K \log \, \Gamma (\alpha_k) \nonumber \\
&& + \sum_{k=1}^K (\alpha_k - 1) \left[ \Psi (\alpha_k) - \Psi \left( \sum_{g=1}^K \alpha_g \right) \right] \\
\sum_{j=1}^{N_i} E_{q_i}[\log \, p(Z_{ij} | \theta_i)] &=& \sum_{j=1}^{N_i} \sum_{k=1}^K \phi_{ijk}
\left[ \Psi (\alpha_k) - \Psi \left( \sum_{g=1}^K \alpha_g \right) \right] \\
\sum_{j=1}^{N_i} E_{q_i}[\log \, p(W_{ij} | Z_{ij}, \tau_{1:K})] &=&
\sum_{j=1}^{N_i} \sum_{k=1}^K \sum_{v=1}^V \phi_{ijk} \, W_{ijv} \, \log \, \tau_{kv} \\
H(q_i) &=& - \Bigg\{ \log \, \Gamma \left( \sum_{k=1}^K \gamma_{ik} \right)
 - \sum_{k=1}^K \log \, \Gamma (\gamma_{ik}) \nonumber \\
 & & + \sum_{k=1}^K (\gamma_{ik} - 1) \left[ \Psi (\gamma_{ik}) - \Psi \left( \sum_{g=1}^K \gamma_{ig} \right) \right] \nonumber \\
 & & + \sum_{j=1}^{N_i} \sum_{k=1}^K \phi_{ijk} \, \log \, \phi_{ijk} \Bigg\}
\end{eqnarray}
\noindent where $\Psi$ denotes the digamma function. All that remains is to derive the fourth term of (\ref{eq:exelbo}):

\begin{eqnarray}
E_q[\log \, p(T_i, \delta_i | Z_{i,1:N_i}, \beta, h_0)] &=& E_q\left[\log \,\left\{ \left[ h_0(T_i) \exp\left( \beta^\prime \bar{Z}_i \right) \right]^{\delta_i}
\times \exp\left[-H_0(T_i) \exp\left( \beta^\prime \bar{Z}_i \right) \right] \right\} \right] \nonumber \\
 &=& E_q\left[\delta_i \, \log \, h_0(T_i) + \delta_i \, \beta^\prime \bar{Z}_i - H_0(T_i) \exp\left( \beta^\prime \bar{Z}_i \right)\right] \nonumber \\
 &=& \delta_i \, \log \, h_0(T_i) + \delta_i E_q\left[\beta^\prime \bar{Z}_i\right] - H_0(T_i) E_q\left[\exp\left( \beta^\prime \bar{Z}_i \right)\right] \nonumber \\
 &=& \delta_i \, \log \, h_0(T_i) + \delta_i \, \beta^\prime \bar{\phi}_i - H_0(T_i)
 \left[ \prod_{j=1}^{N_i} \left( \exp(\frac{\beta}{N_i})^\prime \phi_{ij} \right) \right]
\end{eqnarray} 
\noindent where the $K$-vector \begin{center} $\bar{\phi}_i = (1/N_i) \sum_{j=1}^{N_i} \phi_{ij}$ \end{center}

\noindent We use block coordinate-ascent variational inference, maximizing (\ref{eq:exelbo}) with respect
to $\gamma_i$ and then each $\phi_{ij}$ in turn. As in sLDA \citep{Blei08}, the terms of (\ref{eq:exelbo}) involving $\gamma_i$
are unchanged from LDA and hence the update for $\gamma_i$ is
\begin{align}
\gamma_i^{new} = \alpha + \sum_{j=1}^{N_i} \phi_{ij} \label{eq:newgamma}
\end{align}

\noindent The update for a given $\phi_{ij}$, however, must be derived anew. We first define
the following quantities:
\begin{equation}
\psi_i = \left[\Psi (\gamma_{i1}) - \Psi \left( \sum_{g=1}^K \gamma_{ig} \right), \, \ldots \, , \Psi (\gamma_{iK}) - \Psi \left( \sum_{g=1}^K \gamma_{ig} \right) \right] \nonumber
\end{equation}
\begin{equation}
\xi_{ij} = \left[\sum_{v=1}^V \mathrm{I}(W_{ij} = v) \, \log \, \tau_{1v}, \, \ldots \, , \sum_{v=1}^V \mathrm{I}(W_{ij} = v) \, \log \, \tau_{Kv} \right] \nonumber
\end{equation}

\noindent and then take the partial derivative of (\ref{eq:exelbo}) with respect to $\phi_{ijk}$:
\begin{eqnarray}
\frac{\partial \mathcal{L}}{\partial \phi_{ijk}} &=& 0 + \psi_{ik} + \xi_{ijk} + [- \log \, \phi_{ijk} - 1] \nonumber \\
& & + \, \delta_i \frac{\beta_k}{N_i} - H_0(T_i) \left(\prod_{m \neq j}
\exp\left(\frac{\beta}{N_i}\right)^\prime \phi_{im}\right) \exp\left(\frac{\beta_k}{N_i}\right) \label{eq:newphincomp}
\end{eqnarray}

\noindent Setting this equal to zero and plugging in $\phi_{ijk}^{new}$ yields:
\begin{eqnarray}
\phi_{ijk}^{new} &\propto& \exp\left[ \psi_{ik} + \xi_{ijk} + \delta_i \frac{\beta_k}{N_i} - H_0(T_i) \left( \prod_{m \neq j}
\exp\left(\frac{\beta}{N_i}\right)^\prime \phi_{im} \right) \exp\left(\frac{\beta_k}{N_i}\right) \right] \label{eq:newphin}
\end{eqnarray}
\noindent where proportionality means that the components of
$\phi_{ij}^{new}$ are evaluated according to (\ref{eq:newphin}) and then normalized
so that their sum is one. Variational inference proceeds by iteratively updating the variational parameters
$\{\gamma_i, \phi_{i,1:N_i}\}$ according to (\ref{eq:newgamma}) and (\ref{eq:newphin}) in order to find a local optimum
for the ELBO, which in turn best approximates the evidence given in (\ref{eq:postdoc}).

\subsection*{Parameter estimation}

We use maximum likelihood estimation based on variational expectation-maximization.
Our data are $\mathcal{D} = \{W_{i,1:N_i}, T_i, \delta_i\}$.

\begin{equation}
\mathcal{L}(\alpha, \tau_{1:K}, \beta, h_0 ; \mathcal{D}) =
\sum_{i=1}^{D} \left\{ E_{q_i}[\log \, p(\theta_i, Z_{i,1:N_i}, W_{i,1:N_i}, T_i, \delta_i)] + H(q_i) \right\}
\end{equation}

\noindent In the expectation step (E-step), we use the variational inference algorithm outlined in the first subsection of
Appendix A to estimate the approximate posterior distribution for each document-response pair. In the maximization step (M-step),
we maximize the corpus-level ELBO with respect to $\pi$, subject to some constraints. First, we take $\alpha$
to be ($\alpha_0$/K, \ldots, $\alpha_0$/K) where $\alpha_0$ is specified a priori. This is not necessary; further structure
could be placed on $\alpha$, ranging from a simple Dirichlet prior as in \cite{Blei03} to more complicated
structures allowing dependence among the documents more complex than simple conditional independence
as in \cite{Teh06}. However we, like Blei and McAuliffe \citeyear{Blei08}, prefer letting $\alpha$ be user-defined,
which is simple and straightforward, yet allows some flexibility in the model specification.

The $\tau_{1:K}$ updates are unchanged from unsupervised LDA \citep{Blei03, Blei08} and are thus
calculated in this manner:

\begin{equation}
\hat{\tau}_{kv}^{new} \propto \sum_{i=1}^D \sum_{j=1}^{N_i} \mathrm{I}(W_{ij} = v) \phi_{ijk}
\end{equation}
\noindent where proportionality means that each $\hat{\tau}_k^{new}$ is normalized to sum to one.

The regression coefficients that comprise $\beta$ and the baseline hazard $h_0$ must be numerically optimized
w.r.t. maximizing the portion of the joint ELBO that depends on them.
Numerical optimization is required as no closed form can be derived
in general for the maximizing choice of $\beta$.
The specific computations this process entails depend on the choice
for $h_0$. For example, when an exponential survival model is chosen, so that $h_0 = \lambda$,
$\beta$ and $\lambda$ are numerically optimized by finding the solutions:

\begin{eqnarray}
(\hat{\beta}^{new}, \lambda^{new}) &=& \argmax \, \mathcal{L}(\beta, \lambda) \nonumber \\
&=& \argmax \, \sum_{i=1}^{D} \left[ \delta_i \, \log \, \lambda + \delta_i \, \beta^\prime \bar{\phi}_i - \lambda T_i \times
\prod_{j=1}^{N_i} \exp\left(\frac{\beta}{N_i}\right)^\prime \phi_{ij} \right]
\end{eqnarray}

Numerical optimization for a Weibull survival model is similar. In contrast, if we use a non-parametric Breslow estimate \citep{Breslow74} for $h_0$, we first update $\beta$ given the current
value for $h_0$:

\begin{equation}
\hat{\beta}^{new} = \argmax \, \mathcal{L}(\beta) = \argmax \, \sum_{i=1}^{D} \left[ \delta_i \, \log \, h_0 + \delta_i \, \beta^\prime \bar{\phi}_i - H_0(T_i) \prod_{j=1}^{N_i} \exp\left(\frac{\beta}{N_i}\right)^\prime \phi_{ij} \right]
\end{equation}

Then, given the updated $\beta = \hat{\beta}^{new}$, the maximum likelihood estimate of the baseline hazard $h_0$
at the $r^{th}$ ordered survival time $t_r$ is given by \cite{Breslow74}:

\begin{equation}
\hat{h}_0^{new}(t_r | \beta) = \frac{m_r}{(t_r - t_{r-1}) \sum_{j \in R_r} \exp\left(\beta^\prime \bar{\phi_j}\right)}
\end{equation}

\noindent where $m_r$ is the number of failures at time $t_r$ and $R_r$ is the set of patients that have not failed or been
censored by time $t_r$. Regardless, once $h_0$ has been updated an estimate of the cumulative
baseline hazard $H_0$ follows immediately.

\subsection*{Prediction for a new document}

Given a new patient with document $w_{1:N}$ and a fitted model $\{\alpha, \tau_{1:K}\}$, the posterior mean
$\bar{Z}_{new} = \bar{Z} | w_{1:N}, \alpha, \tau_{1:K}$ can be obtained in order
to estimate from what topics this new patient draws words, and in what proportions.
This is similar to the procedure outlined in the \emph{Posterior Inference} subsection,
except that all survival-related terms in the ELBO are dropped.
Thus, under the same variational distribution as given in (\ref{eq:vardist}), the coordinate
ascent updates are

\begin{eqnarray}
\gamma^{new} &=& \alpha + \sum_{j=1}^N \phi_j \\
\phi_{jk}^{new} &\propto& \exp\left( \psi_ik + \xi_{jk} \right)
\end{eqnarray}
\noindent where again $j$ indexes words, $k$ indexes topics and proportionality
means that the components of $\phi_j^{new}$ are evaluated according to the above update and then normalized
so that their sum is one. Note that this variational sequence is identical to that in \cite{Blei08}, as
they point out that it does not depend on the particular response type.

Given $\bar{Z}_{new}$, measures related to topic membership can be predicted for a new document.
This may be done qualitatively or quantitatively using the chosen survival model.
For example, the predicted median lifetime
can be obtained by solving the following equation for $\widehat{t_{med}}$:

\begin{equation}
\exp\left[ - H_0\left(\widehat{t_{med}}\right) \, \exp\left(\beta^\prime \bar{Z}_{new}\right) \right]= \frac{1}{2}
\end{equation}

\noindent where $H_0$ and $\beta$ are taken from the fitted survLDA model.

\section*{Appendix B: Imposing a background topic}

As mentioned in the text, the documents we are working on are of our own creation and hence we know their nature.
In particular, we know that each patient in the TCGA ovarian cohort received platinum and taxane as treatment
with each of these following surgery was an inclusion criteria. Consequently an uninteresting 
background document would include platinum and taxane
and nothing else. The distribution on the background topic 
may thus be found by placing those weights on those words and re-normalizing so that their sum is one. In order to
avoid negative infinities on the log scale, a small weight such 
as a thousandth of a word may be given to each of the words
that do not appear in this background document. 

With respect to the variational EM algorithm, imposition of a background topic into LDA may be achieved
by specifying one (say the $K^{th}$) topic as the background. In the variational EM, inference in the E-step proceeds
as usual but the M-step is modified so that $\tau_K$ is not updated like the other $\tau_{1:(K-1)}$ but is instead fixed
according to the distribution designated by the user. Modification of survLDA so that it accepts a background topic
is similar except that, in addition to the fixation just outlined, $\beta_K$ is set to $0$. In this way the background
topic also becomes the topic corresponding to the baseline hazard, and the $\beta_k$ values of the other topics will
reflect increased or decreased force of mortality relative to that baseline.

\section*{Acknowledgements}
The authors wish to thank Michael Newton, Michael Jordan, and Kevin Eng for conversations  
that helped to improve the manuscript.

\pagebreak

\bibliographystyle{imsart-nameyear}
\bibliography{survLDABib}

\end{document}